\documentclass[aps,prl,twocolumn,longbibliography]{revtex4-1}
\usepackage{bbm}
\usepackage{graphicx}
\usepackage{dcolumn}
\usepackage{bm}
\usepackage{subfigure}
\usepackage{amsmath}
\usepackage{feynmf}
\usepackage{hyperref}
\usepackage{CJK}
 \usepackage{amssymb}

\usepackage{attachfile}

\newcommand{\bk}{\boldsymbol k}

\newcommand{\bd}{\boldsymbol d}

\newcommand{\bj}{\boldsymbol j}

\usepackage{braket}
\usepackage{esint}
\usepackage{times}

\begin{document}

\title{Extrinsic and Intrinsic Nonlinear Hall Effects across Berry-Dipole Transitions}

\author{Zheng-Yang Zhuang}
\affiliation{Guangdong Provincial Key Laboratory of Magnetoelectric Physics and Devices,
	School of Physics, Sun Yat-Sen University, Guangzhou 510275, China}

\author{Zhongbo Yan}
\email{yanzhb5@mail.sysu.edu.cn}
\affiliation{Guangdong Provincial Key Laboratory of Magnetoelectric Physics and Devices, School of Physics, Sun Yat-Sen University, Guangzhou 510275, China}

\date{\today}

\begin{abstract}
Three-dimensional Hopf insulators are a class of topological phases beyond the tenfold-way classification.
The critical point separating two rotation-invariant Hopf insulator phases with distinct Hopf invariants
is quite different from the usual Dirac-type
or Weyl-type critical points and uniquely characterized by a quantized Berry dipole.
Close to such Berry-dipole transitions, we find that the extrinsic and intrinsic nonlinear Hall conductivity tensors
in the weakly doped regime are characterized by two universal functions of the ratio between
doping level and bulk energy gap, and are directly proportional to
the change in Hopf invariant across the transition. Our work suggests that
the nonlinear Hall effects display a general-sense quantized behavior across
Berry-dipole transitions, establishing a correspondence between
nonlinear Hall effects and Hopf invariant.
\end{abstract}

\maketitle

Quantum responses that can extract topological invariant information
are of great interest in condensed matter physics.
For topological insulating systems, since the topological protection
and the existence of a bulk energy gap can tolerate perturbations,
there may exist quantum responses that are quantized and directly connected to the topological invariant
encoded in the band structure~\cite{Qi2008TFT},
with the quantum (anomalous) Hall insulators being the most well-known example where
the Hall conductance is quantized and connected to the Chern number~\cite{klitzing1980,thouless1982,Chang2013QAHE}.
In comparison, metallic systems are known to hardly support quantized responses since
the Fermi surface deforms under perturbations
and generally lacks of a topological characterization.
Known exceptions include ballistic conductors
where the conductance is shown
to be quantized and connected to the Euler characteristic of the Fermi sea~\cite{vanwess1988,Kane2022},
and Weyl semimetals without inversion and mirror symmetries where
a quantzied circular photogalvanic effect can emerge
in the absence of disorder and interactions~\cite{deJuan2017,Avdoshkin2020}. It is worth mentioning
that the circular photogalvanic effect is a second-order
optical response effect and its quantization in Weyl semimetals is rooted in
the quantized Berry-monopole charge of the Weyl point enclosed
by the Fermi surface~\cite{Armitage2018RMP}.

Recently, extrinsic nonlinear Hall effect (ENHE) and intrinsic nonlinear Hall effect (INHE) in metallic systems,
as another two kinds of second-order quantum responses derived by semiclassical equations
of motion, have gained considerable interest since they have a quantum geometry origin
and can emerge in systems without linear-order anomalous Hall effect~\cite{Ma2021review,Du2021review,Ortix2021}.
It was shown that the ENHE and INHE depend on
the Berry dipole~\cite{Sodemann2015BCD} and Berry-connection polarizability (BCP)~\cite{Gao2014INHE}, respectively,
and both of them are Fermi-surface properties.
The adjectives ``extrinsic'' and ``intrinsic'' applied to distinguish them
reflect one essential difference between these two kinds of nonlinear Hall effects, namely,
the former depends linearly on the relaxation time associated with carrier scatterings~\cite{Sodemann2015BCD},
whereas the latter does not involve any time scale and only depends on the band geometry quantity~\cite{Gao2014INHE}.
Although both of them require the breaking of inversion symmetry, the ENHE is a time-reversal-even effect
and can emerge in systems with time-reversal symmetry,
but the INHE is a time-reversal-odd effect and can only appear
in systems without time-reversal symmetry. Interestingly,
when the system has neither time-reversal symmetry nor inversion symmetry, but their
combination, the $PT$ symmetry, the ENHE is absent as the Berry curvature is forced to vanish~\cite{Xiao2010BP},
whereas the INHE can be significant.  Two recent works have shown  that the INHE could be applied to measure
the N\'{e}el vector of $PT$-symmetric antiferromagnets~\cite{Wang2021INHE,Liu2021INHE}.

Since various quantum geometry quantities
are prominent near band crossings or avoid
crossings, topological semimetals or doped small-gap topological insulators are ideal
material systems to seek for strong ENHE and INHE~\cite{Xu2018BCD,Zhang2018BCDweyl,
	You2018BCD,Zhang2018BCD,Du2018NHE,Facio2018,Ma2019NHE,Kang2019,Battilomo2019BCD,Wang2019NHE,Xiao2019NHE,
	Rostami2020,Sinh2020NHE,Zeng2020BCD,Pantaleon2020BCD,
	Satyam2020BCD,Kumar2021NHE,Liao2021NHE,Zhang2022NHE}. For topological insulators, it is known that
the change of global topology across a topological phase transition is
a result of the dramatic change in quantum geometry near the band edge. Thus,
although the ENHE and INHE are Fermi-surface properties that are generally not possible
to determine the topological invariant associated with the whole Brillouin zone,
they are possible to detect the change of band topology if the topological insulators
are weakly doped.  Previous works indeed found that the ENHE can manifest topological
phase transitions described by low-energy Dirac Hamiltonians
through the dramatic sign change of nonlinear Hall conductivity tensor (NHCT)~\cite{Du2018NHE,Sinha2022,Chakraborty2022}.
However, since therein the tilt or warping of massive Dirac cone is necessary to have nonzero ENHE~\cite{Sodemann2015BCD,Battilomo2019BCD},
the NHCT does not have a simple and universal form
revealing the change of topological invariant.

In this work, we explore the behaviors of ENHE and INHE across
the critical regime of three-dimensional rotation-invariant Hopf insulators.
As a class of topological insulators beyond the tenfold-way classification~\cite{Schnyder2008,kitaev2009periodic},
the Hopf insulators follow a $Z$ classification and do not have time-reversal
and inversion symmetries\cite{Moore2008hopf,deng2013hopf,Liu2017hopf,Alexandradinata2021hopf,Zhu2021hopf}. In addition, the critical points separating two rotation-invariant
Hopf insulator phases with distinct Hopf invariants is characterized by a quantized Berry dipole~\cite{Nelson2021hopf},
which is quite different from the usual Dirac-type or Weyl-type
critical point characterized by a $Z_{2}$ or Berry-monopole charge~\cite{Chiu2015RMP}.
Because of the unique property of the critical point,
we find that close to the transition,
the extrinsic and intrinsic NHCTs in the weakly doped regime
are characterized by two universal functions and explicitly contain the information
of the change in Hopf invariant across the transition.
Although extrinsic and intrinsic NHCTs are independent
quantities of different origins, here they display similar laws and
are simply related across the critical regime.

{\it Rotation-invariant Hopf insulators and Berry-dipole transitions.---}
Hopf insulators have a diversity of model realizations~\cite{Moore2008hopf,deng2013hopf,Liu2017hopf,Neupert2012,Graf2022hopf}.
In this work we focus on the two-band realization for simplicity.
The minimal models for Hopf insulators are constructed by
the so-called Hopf map~\cite{Moore2008hopf}
\begin{eqnarray}
	\mathcal{H}(\bk)=\bd(\bk)\cdot\boldsymbol{\sigma}
\end{eqnarray}
with the three components of the $\bd$-vector given by
\begin{eqnarray}
	d_{i}(\bk)&=&\zeta^{\dag}(\bk)\sigma_{i}\zeta(\bk), \zeta(\bk)=(\zeta_{1}(\bk), \zeta_{2}(\bk))^{T}, \nonumber\\
	\zeta_{1}(\bk)&=&\eta_{1}(\bk)+i\eta_{2}(\bk), \zeta_{2}(\bk)=\eta_{3}(\bk)+i\eta_{4}(\bk),
\end{eqnarray}
where $\sigma_{i}$ are the Pauli matrices, and $\eta_{i}(\bk)$ are real functions
of momentum.
Mathematically, the above equations define a map from $S^{3}$ to $S^{2}$.
In this paper, we focus on models
with rotation symmetry about the $z$ axis. To be specific,
we consider~\cite{Nelson2021hopf}
\begin{eqnarray}
	\zeta_{1}(\bk)&=&\lambda^{n}[\sin (k_{x}a)+i\sin (k_{y}a)]^{n},\nonumber\\
	\zeta_{2}(\bk)&=&\lambda_{z}\sin (k_{z}c)+i[M+t\sum_{i=x,y,z}\cos (k_{i}a_{i})],
\end{eqnarray}
where the lattice constants $a_{x}=a_{y}=a$ and $a_{z}=c$.
The energy spectra of the Hamiltonian take the simple form
\begin{eqnarray}
	E_{\pm}(\bk)=\pm(|\zeta_{1}(\bk)|^{2}+|\zeta_{2}(\bk)|^{2}). \label{spectrum}
\end{eqnarray}
Obviously, the energy gap can only close at time-reversal invariant momenta of
the Brillouin zone. Without loss of generality, we consider that
the band edge is located at $\boldsymbol{\Gamma}=(0,0,0)$,
then an expansion of the complex spinor up to the leading order in momentum
gives
\begin{eqnarray}
	\zeta_{1}(\bk)&=&v^{n}(k_{x}+ik_{y})^{n},\nonumber\\
	\zeta_{2}(\bk)&=&v_{z}k_{z}+im,
\end{eqnarray}
where $v=\lambda a$, $v_{z}=\lambda_{z}c$ and $m=M+3t$.
Accordingly, the low-energy Hopf Hamiltonian has the form
\begin{eqnarray}
	\mathcal{H}(\bk)&=&2v^{n}k_{\rho}^{n}(v_{z}k_{z}\cos n\theta+m\sin n\theta)\sigma_{x}\nonumber\\
	&&+2v^{n}k_{\rho}^{n}(m\cos n\theta-v_{z}k_{z}\sin n\theta)\sigma_{y}\nonumber\\
	&&+(v^{2n}k_{\rho}^{2n}-v_{z}^{2}k_{z}^{2}-m^{2})\sigma_{z},\label{low}
\end{eqnarray}
where $k_{\rho}=\sqrt{k_{x}^{2}+k_{y}^{2}}$, and $\theta$ is the polar angle in the
$k_{x}$-$k_{y}$ plane.
The low-energy spectra are given by
\begin{eqnarray}
	E_{\pm}(\bk)=\pm(v^{2n}k_{\rho}^{2n}+v_{z}^{2}k_{z}^{2}+m^{2}).
\end{eqnarray}
For the simplest case $n=1$, one sees that the low-energy Hamiltonian
at the critical point $m=0$ is distinct to the Weyl Hamiltonian and
the energy dispersion is quadratic rather than the linear dispersion
of usual critical points.

For the Hamiltonian constructed by Hopf map, the Hopf invariant characterizing it
can be determined by~\cite{Liu2017hopf}
\begin{eqnarray}
	N_{h}=\frac{1}{2\pi^{2}}\int d^{3}k \epsilon_{abcd}\hat{\eta}_{a}\partial_{k_{x}}\hat{\eta}_{b}\partial_{k_{y}}\hat{\eta}_{c}
	\partial_{k_{z}}\hat{\eta}_{d},
\end{eqnarray}
where $\hat{\eta}_{i}=\eta_{i}/\sqrt{\eta_{1}^{2}+\eta_{2}^{2}+\eta_{3}^{2}+\eta_{4}^{2}}$,
$\epsilon_{abcd}$ is the fourth-order antisymmetric tensor, and a sum over repeated
indices is implied.
For the low-energy Hopf Hamiltonian, one has~\cite{supplemental}
\begin{eqnarray}
	N_{h}=-\text{sgn}(v_{z}m)\frac{n}{2}.\label{hopf}
\end{eqnarray}
When $m$ changes sign, the Hopf invariant  changes by $n$. It
is worth emphasizing that the low-energy Hopf Hamiltonian can only
determine the change of Hopf invariant since it only faithfully
captures the low-energy part of physics. One needs
the tight-binding Hamiltonian to determine the absolute
values of Hopf invariant before and after the transition.

For the low-energy Hopf Hamiltonian (\ref{low}), the three components
of the Berry curvature can be determined by~\cite{Qi2006qshe}
\begin{eqnarray}
	\Omega_{l}^{(\pm)}(\bk)=\pm\epsilon_{ijl}\frac{\bd(\bk)\cdot(\partial_{k_{i}}\bd(\bk)\times \partial_{k_{j}}\bd(\bk)) }{4d^{3}(\bk)},
\end{eqnarray}
where $d(\bk)=|\bd(\bk)|$ is the norm of the $\bd$-vector, the superscript $+(-)$ refers to the conduction (valence) band,
and $\epsilon_{ijl}$ is the third-order antisymmetric tensor.
An explicit calculation gives~\cite{supplemental}
\begin{eqnarray}
	\Omega_{x}^{(\pm)}(\bk)&=&\pm\frac{2nv_{z}v^{2n}k_{\rho}^{2n-1}(m\sin \theta+v_{z}k_{z}\cos \theta)}{d^{2}(\bk)},\nonumber\\
	\Omega_{y}^{(\pm)}(\bk)&=&\pm\frac{2nv_{z}v^{2n}k_{\rho}^{2n-1}(-m\cos \theta+v_{z}k_{z}\sin \theta)}{d^{2}(\bk)},\nonumber\\
	\Omega_{z}^{(\pm)}(\bk)&=&\pm\frac{2n^{2}v^{2n}k_{\rho}^{2(n-1)}(m^{2}+v_{z}^{2}k_{z}^{2})}{d^{2}(\bk)}.\label{Berry}
\end{eqnarray}
At the critical point $m=0$, it turns out that
the integral of the Berry curvature over a closed surface in momentum space
identically vanishes, however, the integral over the upper ($k_{z}>0$) or lower ($k_{z}<0$)
half of the surface is quantized, namely~\cite{Nelson2021hopf},
\begin{eqnarray}
	\frac{1}{2\pi}\int_{k_{z}>0}\boldsymbol{\Omega}\cdot d\mathbf{S}=n.\label{quantization}
\end{eqnarray}
This unique property again reveals that the critical point between two distinct
Hopf insulator phases is distinct to the Weyl point characterized by
a Berry monopole~\cite{Armitage2018RMP}. Pictorially, the
Berry-flux distribution associated with the critical point
resembles a dipole with the Berry monopole and antimonopole forming the dipole infinitely close
and mirror-symmetric about the $k_{z}=0$ plane~\cite{Nelson2021hopf,Sun2018mirrorweyl}.
The results in Eqs.{\ref{hopf} and (\ref{quantization}) indicate
	that the change of Hopf invariant across the transition is equal to the quantized
	value of the Berry dipole at the critical point, similar to
	that the change of Chern number in a two-dimensional system
	is equal to the Berry-monopole charge it crosses~\cite{Armitage2018RMP}.

	Since the critical point of rotation-invariant Hopf insulators is characterized by a quantized Berry dipole
	and the ENHE is related to the Berry dipole,
	it is natural to expect that the ENHE would display some novel behaviors
	across such Berry-dipole transitions.

	{\it ENHE across Berry-dipole transitions.---} Sodemann and Fu revealed
	that an ac electric field would result in a mixture of dc and second-harmonic
	Hall-type currents in the second-order-response regime in systems without inversion
	symmetry\cite{Sodemann2015BCD}, i.e., $\bj=\bj^{0}+\bj^{2\omega}$ with
	$j^{0}_{i}=\chi_{ijk}\mathcal{E}_{j}\mathcal{E}_{k}^{*}$ and $j^{2\omega}_{i}=\chi_{ijk}\mathcal{E}_{j}\mathcal{E}_{k}$.
	The extrinsic NHCT is given by\cite{Sodemann2015BCD}
	\begin{eqnarray}
		\chi_{ijk}^{\rm ext}=-\frac{e^{3}\tau}{2(1+i\omega \tau)}\epsilon_{ilk}D_{jl}, \label{coeff}
	\end{eqnarray}
	where $\tau$ denotes the relaxation time whose value depends on the carrier scattering.
	$D_{jl}$ is the Berry dipole given by~\cite{Sodemann2015BCD}
	\begin{eqnarray}
		D_{jl}=-\sum_{\alpha}\int\frac{d^{3}k}{(2\pi)^{3}}\partial_{k_{j}}f^{(\alpha)}(\bk)\Omega_{l}^{(\alpha)}(\bk),\label{bcd}
	\end{eqnarray}
	where the sum is over all bands. The derivative of the Fermi-Dirac distribution function, $\partial_{k_{j}}f^{(\alpha)}$,
	is equal to $-\partial_{k_{j}}E_{\alpha}\delta(\mu-E_{\alpha})$
	at the zero-temperature limit, indicating that the Berry dipole $D_{jl}$ is
	a Fermi-surface property.
	
	Using the Berry curvature in Eq.(\ref{Berry}), we find that the extrinsic NHTC
	only contains four nonzero components, including
	$\chi_{zxx}^{\rm ext}$, $\chi_{xxz}^{\rm ext}$, $\chi_{zyy}^{\rm ext}$, $\chi_{yyz}^{\rm ext}$.
	However, there is in fact only one independent component. This can be inferred by noting that
	the rotation symmetry forces $\chi_{zxx}^{\rm ext}=\chi_{zyy}^{\rm ext}$, and the Hall nature
	forces $\chi_{zxx}^{\rm ext}=-\chi_{xxz}^{\rm ext}$ and $\chi_{zyy}^{\rm ext}=-\chi_{yyz}^{\rm ext}$,
	which can also be simply inferred from the antisymmetric property of the Levi-Civita symbol
	in Eq.(\ref{coeff}). Focusing on
	the zero-temperature limit and only showing the explicit form of $\chi_{zxx}^{\rm ext}$, one has~\cite{supplemental}
	\begin{eqnarray}
		\chi_{zxx}^{\rm ext}=N_{h}\chi_{1}\mathcal{B}_{1}(\mu/m^{2}),\label{ENLE}
	\end{eqnarray}
	where $N_{h}$ is the Hopf invariant given by Eq.(\ref{hopf}),
	$\chi_{1}=e^{3}\tau/2(1+i\omega \tau)$, and
	$\mathcal{B}_{1}(x)$ is a dimensionless universal function of the form
	\begin{eqnarray}
		\mathcal{B}_{1}(x)=\frac{2(x-1)^{3/2}}{3\pi^2 x^2}\Theta(x-1).
	\end{eqnarray}
	Here electronic doping is assumed, i.e., $\mu>0$. Since the energy gap of the low-energy Hamiltonian (\ref{low})
	is equal to $2m^{2}$, $\mathcal{B}_{1}(\mu/m^{2})$ is a universal function of the ratio between chemical
	potential and bulk energy gap.
	
	According to Eqs.(\ref{hopf}) and (\ref{ENLE}), one sees
	that the extrinsic NHCT reverses its sign
	across a Berry-dipole transition, thus allowing a sensitive probe of the
	transition. Remarkably, if the relaxation time is assumed to be constant,
	the extrinsic NHCT turns out to be a universal function multiplied
	with the change of Hopf invariant across the transition.
	To the best of our knowledge, this is the thus-far-only-known system that the ENHE
	displays a general-sense quantized behavior with
	a correspondence to topological invariant.
	Using this result, it is possible to precisely probe
	the change in topological invariant
	across Berry-dipole transitions by measuring the evolution of current
	with the change of doping level.

	{\it INHE across Berry-dipole transitions.---} Since
	the time-reversal and inversion symmetries are both
	broken in the Hopf Hamiltonian, the INHE is also
	allowed. However, quite
	different from the ENHE, the INHE introduced by
	Gao, Yang and Niu depends on a band geometry quantity
	known as BCP~\cite{Gao2014INHE}, which does not
	have a direct connection with the Berry curvature.
	Although the ENHE and INHE have different origins,
	as we will show below, a simple relation exists between the Berry curvature and the BCP for the low-energy
	Hopf Hamiltonian, leading to that the INHE displays a similar behavior like the ENHE
	across Berry-dipole transitions.

	The Hall current originating from INHE has the form $j_{\alpha}^{\rm int}
	=\chi_{\alpha\beta\gamma}^{\rm int}\mathcal{E}_{\beta}\mathcal{E}_{\gamma}$, where the
	intrinsic NHCT is given by~\cite{Gao2014INHE,Wang2021INHE,Liu2021INHE}
	\begin{eqnarray}
		\chi_{\alpha\beta\gamma}^{\rm int}=-e^{3}\sum_{n}\int \frac{d^{3}k}{(2\pi)^{3}}v_{\alpha}^{(n)}G_{\beta\gamma}^{(n)}
		\frac{\partial f(E_{n})}{\partial E_{n}}-(\alpha\leftrightarrow\beta),\quad\quad\label{INHE}
	\end{eqnarray}
	where $E_{n}$, $v_{\alpha}^{(n)}=\partial E_{n}/\partial k_{\alpha}$ and $G_{\beta\gamma}^{(n)}$
	denote the $n$th band's dispersion, velocity in the $\alpha$ direction
	and BCP, respectively.  Their $\bk$-dependence are made implicit in Eq.(\ref{INHE}).
	The gauge-independent BCP is given by~\cite{Gao2014INHE,Wang2021INHE,Liu2021INHE}
	\begin{eqnarray}
		G^{(n)}_{\beta\gamma}(\bk)=2\text{Re}\sum_{m\neq n}\frac{A_{nm,\beta}(\bk)A_{mn,\gamma}(\bk)}{E_{n}(\bk)-E_{m}(\bk)},
	\end{eqnarray}
	where $A_{nm,\beta}(\bk)=i\langle u_{n}(\bk)|\partial_{k_{\beta}}u_{m}(\bk)\rangle$ is the interband Berry connection.
	From Eq.(\ref{INHE}), it is apparent that
	$\chi_{\alpha\beta\gamma}^{\rm int}=-\chi_{\beta\alpha\gamma}^{\rm int}$.

	For the two-band Hamiltonian considered, one has~\cite{Gao2014INHE}
	\begin{eqnarray}
		G^{(+)}_{\beta\gamma}(\bk)=\frac{\partial_{k_{\beta}}\hat{\bd}(\bk)\cdot\partial_{k_{\gamma}}
			\hat{\bd}(\bk)}{4d(\bk)}=-G^{(-)}_{\beta\gamma}(\bk),\label{BCP}
	\end{eqnarray}
	where $\hat{\bd}(\bk)=\bd(\bk)/d(\bk)$ is the normalized $\bd$-vector. Interestingly,
	the numerator in Eq.(\ref{BCP}) suggests that the BCP is related to the quantum metric
	for a two-band Hamiltonian. Since the
	BCP is a symmetric tensor, i.e., $G^{(\pm)}_{\beta\gamma}(\bk)=G^{(\pm)}_{\gamma\beta}(\bk)$,
	there are six independent components.
	By straightforward calculations, one has~\cite{supplemental}
	\begin{eqnarray}
		G_{xx}^{(\pm)}(\bk)&=&G_{yy}^{(\pm)}(\bk)=\pm\frac{n^2v^{2n}k_{\rho}^{2n-2}(m^{2}+v_{z}^{2}k_{z}^{2})}{d^{3}(\bk)},\nonumber\\
		G_{zz}^{(\pm)}(\bk)&=&\pm\frac{v_{z}^{2}v^{2n}k_{\rho}^{2n}}{d^{3}(\bk)},\quad G_{xy}^{(\pm)}(\bk)=0,\nonumber\\
		G_{xz}^{(\pm)}(\bk)&=&\mp\frac{nv^{2n}v_{z}k_{\rho}^{2n-1}(m\sin \theta+v_{z}k_{z}\cos \theta)}{d^{3}(\bk)},\nonumber\\
		G_{yz}^{(\pm)}(\bk)&=&\mp\frac{nv^{2n}v_{z}k_{\rho}^{2n-1}(-m\cos \theta+v_{z}k_{z}\sin \theta)}{d^{3}(\bk)}.\label{metric}
	\end{eqnarray}
	A close look of Eqs.(\ref{Berry}) and (\ref{metric}) finds that four of the six independent
	components of the BCP tensor have a simple relation with
	the three components of the Berry curvature, i.e.,
	\begin{eqnarray}
		\Omega_{x}^{(\pm)}=-2dG_{xz}^{(\pm)}, \Omega_{y}^{(\pm)}=-2dG_{yz}^{(\pm)},  \Omega_{z}^{(\pm)}=2dG_{xx(yy)}^{(\pm)}.\quad
	\end{eqnarray}
	Bringing Eq.(\ref{metric}) into Eq.(\ref{INHE}), one finds that $\chi_{\alpha\beta\gamma}^{\rm int}$ is nonzero only
	when $\alpha\neq\beta\neq\gamma$, and the nonzero independent components have the simple relation~\cite{supplemental}
	\begin{eqnarray}
		\chi_{xyz}^{\rm int}=-2\chi_{yzx}^{\rm int}=-2\chi_{zxy}^{\rm int}.
	\end{eqnarray}
	Also focusing on
	the zero-temperature limit and only showing the explicit form of $\chi_{xyz}^{\rm int}$, one has~\cite{supplemental}
	\begin{eqnarray}
		\chi_{xyz}^{\rm int}=N_{h}\chi_{2}\mathcal{B}_{2}(\mu/m^{2}),\label{INHEcomponent}
	\end{eqnarray}
	where $\chi_{2}=e^{3}/m^{2}$ and
	\begin{eqnarray}
		\mathcal{B}_{2}(x)=\frac{2(x-1)^{3/2}}{3\pi^{2}x^{3}}\Theta(x-1)=\frac{\mathcal{B}_{1}(x)}{x}.
	\end{eqnarray}
	
	Compared to the ENHE, one sees that the INHE displays a similar behavior, but
	a big difference is that the factor $\chi_{2}$ suggests that the peak of the intrinsic
	NHCT goes divergent as the bulk energy gap decreases
	to infinitely small but nonzero (note that the critical point of the low-energy
	Hopf Hamiltonian has emergent inversion symmetry so the INHE is forced to vanish
	when $m=0$). This property implies that the INHE can provide an even more sensitive
	probe of the Berry-dipole transition. Furthermore,
	one can find $\chi^{\rm ext}/\chi^{\rm int}\sim m^{2}\tau$, thus a
	simultaneous study of ENHE and INHE can also provide a probe
	of the relaxation time.

	{\it Discussions and conclusions.---} It is worth emphasizing that
	because the low-energy Hopf Hamiltonian does not have time-reversal symmetry,
	the linear Hall conductivity tensor $\sigma_{ij}$ does not identically vanish. By
	analyzing the Berry curvature in Eq.(\ref{Berry}), it is easy to find that $\sigma_{xz}$
	and $\sigma_{yz}$ identically vanish, and only $\sigma_{xy}$ is finite in the doped regime.
	This can also be figured out by noting that, despite the absence of global time-reversal symmetry,
	the Hamiltonian has spinless time-reversal
	symmetry at any $k_{y}$ or $k_{x}$ plane. This result suggests that
	the linear anomalous Hall effect only appears in the $xy$ plane. For
	the ENHE and INHE, their nonzero components indicate that the Hall current
	will flow in the $z$ direction if the electric vector lies in the $xy$ plane,
	thus they can be easily distinguished from the linear-order Hall current according to
	the current direction. Of course, they can also be distinguished by using lock-in
	method since the linear-order and second-order Hall signals have different dependence
	on the frequency of the ac electric field.
	The ENHE and INHE can also be easily distinguished from each other
	by controlling the electric vector to lie in the $xz$ or $yz$ plane.
	For the former, the current will flow in parallel with the electric component
	in the $xy$ plane, while the latter will flow perpendicular to the electric vector plane.
	
	Now we discuss potential systems to observe our predictions.
	In theory, the two-band Hopf insulators considered provide the simplest
	realization of Berry-dipole transitions, however, two-band Hopf insulators
	remain elusive in experiments. Since the critical point associated with a quantized Berry dipole
	is equivalent to the overlap of two mirror-symmetry-related Weyl points~\cite{Sun2018mirrorweyl}, a potential
	route to realize the Berry-dipole transition is to consider
	magnetic Weyl semimetals with a mirror plane and then break appropriate  symmetries to
	gap out the nodal points~\cite{Nelson2021hopf}. The experimental implementation
	of two-band Hopf insulators has also been explored in the context of
	cold-atom systems, and it was suggested that cold-atom systems with
	long-range dipolar interaction and Floquet modulations may realize
	the long-sought Hopf insulators~\cite{Schuster2021a,Schuster2021b}. On the other hand, a recent work
	revealed that
	Bloch oscillations provide a counterpart realization of ENHE in
	cold-atom systems~\cite{Chen2022NHE}. Therefore, the cold-atom systems are a potential
	platform to observe our predictions. Furthermore, it is worth pointing
	out that Berry-dipole transitions are not restricted to two-band realizations.
	There also exist $N$-fold ($N\geq3$)
	Berry-dipole critical points with linear dispersion~\cite{Graf2022hopf}. Compared to the
	two-band realizations of Berry-dipole transitions, an important
	advantage of the $N$-band ($N\geq3$) realizations is that their
	corresponding tight-binding Hamiltonians can only involve nearest-neighbor
	hoppings, and are thus more feasible in experiments.
	By investigating the $N=3$ case, we find that the ENHE and INHE
	do also display the expected general-sense quantized behaviors across
	the Berry-dipole transition~\cite{supplemental}. The only different is that the
	associated universal functions have a different form
	due to the difference in dispersion.
	This suggests that this class of systems can also be applied
	to test our predictions on the ENHE and INHE across Berry-dipole
	transitions.

	In conclusion, we have shown that the nonlinear Hall effects
	display a general-sense quantized behavior across Berry-dipole transitions,
	building up a correspondence between nonlinear Hall effects
	and Hopf invariant. A direction for future study is
	to consider semimetals with nodal points or insulators
	with critical points  characterized by higher-order Berry moments
	and explore what kinds of quantum responses can uniquely reflect
	their existence.

	{\it Acknowledgements.---} The authors would like to
	thank Prof. Qian Niu, Prof. Yang Gao and Prof. Zhi Wang for
	illuminating discussions. This work is supported by the National Natural Science Foundation of China (Grant No.11904417
	and 12174455) and the Natural Science Foundation of Guangdong Province
	(Grant No. 2021B1515020026).

\bibliography{dirac}

\begin{widetext}
\clearpage
\begin{center}
\textbf{\large Supplemental Material for \\``Extrinsic and Intrinsic Nonlinear Hall Effects across Berry-Dipole Transitions''}\\
\vspace{4mm}
{Zheng-Yang Zhuang and Zhongbo Yan$^{*}$}\\
\vspace{2mm}
{\em Guangdong Provincial Key Laboratory of Magnetoelectric Physics and Devices, School of Physics, \\Sun Yat-Sen University, Guangzhou 510275, China}\\
\end{center}

\setcounter{equation}{0}
\setcounter{figure}{0}
\setcounter{table}{0}
\makeatletter
\renewcommand{\theequation}{S\arabic{equation}}
\renewcommand{\thefigure}{S\arabic{figure}}
\renewcommand{\bibnumfmt}[1]{[S#1]}

The supplemental material contains the details for the
derivation of the extrinsic and intrinsic nonlinear Hall
conductivity tensors.
Six sections are in order: (I) Hopf invariant of the
low-energy Hopf Hamiltonian.
(II) Berry curvature for the
low-energy Hopf Hamiltonian.
(III) Berry connection polarizability for the low-energy Hopf Hamiltonian.
(IV) Extrinsic nonlinear Hall tensor for the low-energy Hopf Hamiltonian.
(V) Intrinsic nonlinear Hall tensor for the low-energy Hopf Hamiltonian.
(VI) Extrinsic and intrinsic Hall conductivity tensors across three-band Berry-dipole transitions.


\section{I. Hopf invariant of the low-energy Hopf Hamiltonian}

As discussed in the main article, the minimal realization of two-band Hopf-insulator Hamiltonians is given by Hopf map. Put it explicitly,
the two-band Hopf Hamiltonian is given by
\begin{eqnarray}
	H(\bk)=\bd(\bk)\cdot\boldsymbol{\sigma},
\end{eqnarray}
where the three components of the $\bd$-vector are determined by the Hopf map~\cite{Moore2008hopf},
\begin{eqnarray}
	d_{i}(\bk)&=&\zeta^{\dag}\sigma_{i}\zeta, \quad \zeta=(\zeta_{1}, \zeta_{2})^{T}, \nonumber\\
	\zeta_{1}&=&\eta_{1}+i\eta_{2}, \zeta_{2}=\eta_{3}+i\eta_{4}, 
\end{eqnarray}
where $\sigma_{i}$ are the Pauli matrices, and $\eta_{i}$ are real functions
of momentum. To guarantee that the Hamiltonian has
rotation symmetry about the $z$ axis and allows topological phase transitions
associated with an arbitrary change of Hopf invariant~\cite{Nelson2021hopf}, we choose
\begin{eqnarray}
	\zeta_{1}&=&\lambda^{n}[\sin (k_{x}a)+i\sin (k_{y}a)]^{n},\nonumber\\
	\zeta_{2}&=&\lambda_{z}\sin (k_{z}c)+i[M+t\sum_{i=x,y,z}\cos (k_{i}a_{i})],
\end{eqnarray}
where the lattice constants $a_{x}=a_{y}=a$ and $a_{z}=c$. As our focus is on 
the critical regime, without loss of generality, we consider that
the band edge is located at $\boldsymbol{\Gamma}=(0,0,0)$,
then an expansion of the complex spinor up to the leading order in momentum
gives
\begin{eqnarray}
	\zeta_{1}&=&v^{n}(k_{x}+ik_{y})^{n},\nonumber\\
	\zeta_{2}&=&v_{z}k_{z}+im,
\end{eqnarray}
where $v=\lambda a$, $v_{z}=\lambda_{z}c$ and $m=M+3t$.
Accordingly, the corresponding three components of the $\bd$-vector have the form
\begin{eqnarray}
	d_{x}(\bk)&=&2v^{n}k_{\rho}^{n}[v_{z}k_{z}\cos(n\theta)+m\sin(n\theta)], \nonumber\\
	d_{y}(\bk)&=&2v^{n}k_{\rho}^{n}[m\cos(n\theta)-v_{z}k_{z}\sin(n\theta)], \nonumber\\
	d_{z}(\bk)&=&v^{2n}k_{\rho}^{2n}-v_{z}^{2}k_{z}^{2}-m^{2},
\end{eqnarray}
and the low-energy Hopf Hamiltonian has the form
\begin{eqnarray}
	\mathcal{H}(\bk)&=&2v^{n}k_{\rho}^{n}[v_{z}k_{z}\cos(n\theta)+m\sin(n\theta)]\sigma_{x}\nonumber\\
	&&+2v^{n}k_{\rho}^{n}[m\cos(n\theta)-v_{z}k_{z}\sin(n\theta)]\sigma_{y}\nonumber\\
	&&+(v^{2n}k_{\rho}^{2n}-v_{z}^{2}k_{z}^{2}-m^{2})\sigma_{z},\label{low}
\end{eqnarray}
where $k_{\rho}=\sqrt{k_{x}^{2}+k_{y}^{2}}$, and $\theta$ is the polar angle.

The energy spectra are given by
\begin{eqnarray}
	E_{\pm}(\bk)=\pm\sqrt{\sum_{i}d_{i}^{2}(\bk)}=\pm d(\bk)=\pm(v^{2n}k_{\rho}^{2n}+v_{z}^{2}k_{z}^{2}+m^{2}),
\end{eqnarray}
where $d(\bk)=(v^{2n}k_{\rho}^{2n}+v_{z}^{2}k_{z}^{2}+m^{2})$ is the norm of the $\bd$-vector.

For the low-energy Hopf Hamiltonian, its topology is characterized by the Hopf invariant~\cite{Liu2017hopf}
\begin{eqnarray}
	N_{h}=\frac{1}{2\pi^{2}}\int d^{3}k \epsilon^{abcd}\hat{\eta}_{a}\partial_{k_{x}}\hat{\eta}_{b}\partial_{k_{b}}\hat{\eta}_{c}
	\partial_{k_{z}}\hat{\eta}_{d}
\end{eqnarray}
where $\boldsymbol{\hat{\eta}}=(\hat{\eta}_{1},\hat{\eta}_{2},\hat{\eta}_{3},\hat{\eta}_{4})=
(v^{n}\text{Re}(k_{x}+ik_{y})^{n},v^{n}\text{Im}(k_{x}+ik_{y})^{n},v_{z}k_{z},m)/\sqrt{v^{2n}k_{\rho}^{2n}+v_{z}^{2}k_{z}^{2}+m^{2}}$. Applying the
above formula, one has
\begin{eqnarray}
	N_{h}&=&-\frac{1}{2\pi^{2}}\int d^{3}k \frac{n^{2}v_{z}mv^{2n}k_{\rho}^{2n-2}}{(v^{2n}k_{\rho}^{2n}+v_{z}^{2}k_{z}^{2}+m^{2})^{2}}\nonumber\\
	&=&-\frac{1}{\pi}\int_{0}^{+\infty}k_{\rho}dk_{\rho}\int_{-\infty}^{\infty}dk_{z}
	\frac{n^{2}v_{z}mv^{2n}k_{\rho}^{2n-2}}{(v^{2n}k_{\rho}^{2n}+v_{z}^{2}k_{z}^{2}+m^{2})^{2}}\nonumber\\
	&=&-\frac{nm}{2\pi}\int_{0}^{+\infty}d(v^{2n}k_{\rho}^{2n})\int_{-\infty}^{\infty}dk_{z}
	\frac{v_{z}}{(v^{2n}k_{\rho}^{2n}+v_{z}^{2}k_{z}^{2}+m^{2})^{2}}\nonumber\\
	&=&-\frac{nm}{2\pi}\int_{0}^{+\infty}dx\int_{-\infty}^{\infty}dk_{z}
	\frac{v_{z}}{(x+v_{z}^{2}k_{z}^{2}+m^{2})^{2}}\nonumber\\
	&=&-\frac{nm}{2\pi}\int_{-\infty}^{\infty}dk_{z}
	\frac{v_{z}}{v_{z}^{2}k_{z}^{2}+m^{2}}\nonumber\\
	&=&-\text{sgn}(v_{z}m)\frac{n}{2}.
\end{eqnarray}
The result indicates that the Hopf invariant will change by $n$ when $m$ changes its sign.

\section{II. Berry curvature for the low-energy Hopf Hamiltonian}

Since the extrinsic nonlinear Hall effect depends on the Berry curvature, in this section we calculate
all components of the Berry curvature. For the two-band model, it is known that the Berry curvature can be determined
by the formula~\cite{Qi2006qshe}
\begin{eqnarray}
	\Omega_{l}^{(\pm)}(\bk)=\pm\epsilon_{ijl}\frac{\bd(\bk)\cdot(\partial_{k_{i}}\bd(\bk)\times \partial_{k_{j}}\bd(\bk)) }{4d^{3}(\bk)},
\end{eqnarray}
where the superscripts $\pm$ refer to the conduction and valence bands, respectively. $\epsilon_{ijl}$ is 
the Levi-Civita symbol which is an antisymmetric third-order tensor,  and here
a sum over repeated indices is implied.
By straightforward calculations, one has
\begin{eqnarray}
	\Omega_{x}^{(\pm)}(\bk)&=&\pm\frac{2nv_{z}v^{2n}k_{\rho}^{2n-1}(m\sin\theta+v_{z}k_{z}\cos\theta)}{d^{2}(\bk)},\nonumber\\
	\Omega_{y}^{(\pm)}(\bk)&=&\pm\frac{2nv_{z}v^{2n}k_{\rho}^{2n-1}(-m\cos\theta+v_{z}k_{z}\sin\theta)}{d^{2}(\bk)},\nonumber\\
	\Omega_{z}^{(\pm)}(\bk)&=&\pm\frac{2n^{2}v^{2n}k_{\rho}^{2(n-1)}(m^{2}+v_{z}^{2}k_{z}^{2})}{d^{2}(\bk)}.\label{SBerry}
\end{eqnarray}
At the critical point $m=0$, the Berry curvature can be written compactly as
\begin{eqnarray}
	\boldsymbol{\Omega}^{(\pm)}(\bk)=\pm\frac{2nv_{z}^{2}v^{2n}k_{\rho}^{2(n-1)}k_{z}}{d^{2}(\bk)}(k_{x},k_{y},nk_{z}).
\end{eqnarray}
For $n=1$, and $v=v_{z}$, one has
\begin{eqnarray}
	\boldsymbol{\Omega}^{(\pm)}(\bk)=\pm\frac{2k_{z}\bk}{k^{4}}.
\end{eqnarray}
Considering a sphere as the integral contour, it is easy to find that 
\begin{eqnarray}
	\oint_{S}\boldsymbol{\Omega}^{(\pm)}\cdot d\mathbf{S}=\pm\int_{0}^{2\pi}d\phi\int_{0}^{\pi}d\theta2\cos\theta\sin\theta=0,
\end{eqnarray} 
but an integral over half of the sphere gives 
\begin{eqnarray}
	\int_{k_{z}>0}\boldsymbol{\Omega}^{(\pm)}\cdot d\mathbf{S}=\pm\int_{0}^{2\pi}d\phi
	\int_{0}^{\frac{\pi}{2}}d\theta2\cos\theta\sin\theta=\pm 2\pi.
\end{eqnarray}

\section{III. Berry connection polarizability for the low-energy Hopf Hamiltonian}

Since the intrinsic nonlinear Hall effect depends on the  Berry connection polarizability (BCP), in this section we calculate
all components of the BCP tensor. For a two-band model, the BCP is given by
\begin{eqnarray}
	G_{\alpha\beta}^{(+)}(\bk)=-2\text{Re}\frac{\langle u^{+}(\bk)|\partial_{k_{\alpha}} u^{-}(\bk)\rangle\langle u^{-}(\bk)|\partial_{k_{\beta}} u^{+}(\bk)\rangle}{E_{+}(\bk)-E_{-}(\bk)},
\end{eqnarray}
where $|u^{\pm}(\bk)\rangle$ refer to the eigenstates for the conduction ($E_{+}(\bk)$) and
valence ($E_{-}(\bk)$) bands, respectively. Using the projection operators $\hat{P}(\bk)=|u^{-}(\bk)\rangle\langle
u^{-}(\bk)|$, $\hat{Q}(\bk)=|u^{+}(\bk)\rangle\langle u^{+}(\bk)|$ and
the orthogonal relation $\langle
u^{+}(\bk)|u^{-}(\bk)\rangle=0$, one has
\begin{eqnarray}
	&\langle u^{+}(\bk)|\partial_{k_{\alpha}} u^{-}(\bk)\rangle\langle u^{-}(\bk)|\partial_{k_{\beta}} u^{+}(\bk)\rangle\nonumber\\
	&=\langle u^{+}(\bk)|\partial_{k_{\alpha}}\hat{P}(\bk)\partial_{k_{\beta}}\hat{Q}(\bk) |u^{+}(\bk)\rangle.
\end{eqnarray}
Furthermore,
\begin{eqnarray}
	&&\text{Re}\langle u^{+}(\bk)|\partial_{k_{\alpha}} u^{-}(\bk)\rangle\langle u^{-}(\bk)|\partial_{k_{\beta}} u^{+}(\bk)\rangle\nonumber\\
	&=&\text{Re}\langle \partial_{k_{\alpha}} u^{-}(\bk) |u^{+}(\bk)\rangle\langle \partial_{k_{\beta}} u^{+}(\bk)|u^{-}(\bk)\rangle\nonumber\\
	&=&\text{Re}\langle u^{-}(\bk)|\partial_{k_{\alpha}} \hat{P}(\bk) \partial_{k_{\beta}} \hat{Q}(\bk)|u^{-}(\bk)\rangle.
\end{eqnarray}
A combination of the two above equations leads to
\begin{eqnarray}
	G_{\alpha\beta}^{(+)}(\bk)=-\frac{\text{Re}\{\text{Tr}[\partial_{k_{\alpha}} \hat{P}(\bk) \partial_{k_{\beta}}\hat{Q}(\bk)]\}}{E_{+}(\bk)-E_{-}(\bk)}.
\end{eqnarray}
For the two-band model,
\begin{eqnarray}
	\hat{P}(\bk)=\frac{I-\hat{\bd}(\bk)\cdot\boldsymbol{\sigma}}{2}, \quad \hat{Q}(\bk)=\frac{I+\hat{\bd}(\bk)\cdot\boldsymbol{\sigma}}{2},
\end{eqnarray}
where $I$ is the two-by-two identity matrix, and $\hat{\bd}(\bk)=\bd(\bk)/d(\bk)$ is the normalized $\bd$-vector.
A straightforward calculation then reveals that
\begin{eqnarray}
	G_{\alpha\beta}^{(+)}(\bk)&=&\frac{\partial_{k_{\alpha}} \hat{\bd}(\bk)\cdot\partial_{k_{\beta}} \hat{\bd}(\bk)}{4d(\bk)}=G_{\beta\alpha}^{(+)}(\bk).\label{metric}
\end{eqnarray}
Similar analysis reveals that $G_{\alpha\beta}^{(-)}(\bk)=-G_{\alpha\beta}^{(+)}(\bk)$. Because
the BCP is a symmetric tensor, it has six independent components.

By applying Eq.(\ref{metric}), we find
\begin{eqnarray}
	G_{xx}^{(\pm)}(\bk)&=&G_{yy}^{(\pm)}(\bk)=\pm\frac{n^2v^{2n}k_{\rho}^{2n-2}(m^{2}+v_{z}^{2}k_{z}^{2})}{d^{3}(\bk)},\nonumber\\
	G_{zz}^{(\pm)}(\bk)&=&\pm\frac{v_{z}^{2}v^{2n}k_{\rho}^{2n}}{d^{3}(\bk)},\nonumber\\
	G_{xy}^{(\pm)}(\bk)&=&G_{yx}^{(\pm)}(\bk)=0,\nonumber\\
	G_{xz}^{(\pm)}(\bk)&=&G_{zx}^{(\pm)}(\bk)=\mp\frac{nv_{z}v^{2n}k_{\rho}^{2n-1}(m\sin\theta+v_{z}k_{z}\cos\theta)}{d^{3}(\bk)},\nonumber\\
	G_{yz}^{(\pm)}(\bk)&=&G_{zy}^{(\pm)}(\bk)=\mp\frac{nv_{z}v^{2n}k_{\rho}^{2n-1}(-m\cos\theta+v_{z}k_{z}\sin\theta)}{d^{3}(\bk)}.\label{Smetric}
\end{eqnarray}
A close look of Eqs.(\ref{SBerry}) and (\ref{Smetric}) reveals that four of
the six independent components of the BCP tensor has a simple relation with
the three components of the Berry curvature, i.e.,
\begin{eqnarray}
	\Omega_{x}^{(\pm)}=-2dG_{xz}^{(\pm)},\, \Omega_{y}^{(\pm)}=-2dG_{yz}^{(\pm)}, \, \Omega_{z}^{(\pm)}=2dG_{xx}^{(\pm)}=2dG_{yy}^{(\pm)}.
\end{eqnarray}

\section{IV. Extrinsic nonlinear Hall tensor for the low-energy Hopf Hamiltonian}

In this section, we show the details for the derivation of the extrinsic nonlinear
Hall conductivity tensor. For notational simplicity, we define
$\chi_{1}=e^{3}\tau/2(1+i\omega \tau)$, then the form of the extrinsic nonlinear Hall conductivity
tensor is given by
\begin{eqnarray}
	\chi_{ijk}^{\rm ext}=-\chi_{1}\epsilon_{ilk}D_{jl},\label{tensor}
\end{eqnarray}
where the so-called Berry curvature dipole is given by~\cite{Sodemann2015BCD}
\begin{eqnarray}
	D_{jl}&=&-\sum_{\alpha}\int\frac{d^{3}k}{(2\pi)^{3}}\partial_{j}f^{(\alpha)}(\bk)\Omega_{l}^{(\alpha)}(\bk)\nonumber\\
	&=&\sum_{\alpha}\int\frac{d^{3}k}{(2\pi)^{3}}f^{(\alpha)}(\bk)\partial_{j}\Omega_{l}^{(\alpha)}(\bk).\label{bcd}
\end{eqnarray}
Here $\alpha$ refers to the band index, and $f^{(\alpha)}=f(E_{\alpha}, \mu, T)$ refers to the corresponding
Fermi-Dirac distribution function at chemical potential $\mu$ and temperature $T$. In the following, we will focus
on the zero-temperature limit. For the two-by-two low-energy Hopf Hamiltonian, $\alpha=\pm$. Because of
the relation $\Omega_{l}^{(+)}(\bk)=-\Omega_{l}^{(-)}(\bk)$, below we will focus on $\mu>0$.

According to the Berry curvature calculated,
we find that there are only four nonzero components, including $\chi_{zyy}^{\rm ext}$,
$\chi_{yyz}^{\rm ext}$, $\chi_{zxx}^{\rm ext}$ and $\chi_{xxz}^{\rm ext}$. In Eq.(\ref{tensor}),
the Levi-Civita symbol indicates that the conductivity tensor
is antisymmetric about the first and third indices. Therefore,
$\chi_{zyy}^{\rm ext}=-\chi_{yyz}^{\rm ext}$, and $\chi_{zxx}^{\rm ext}=-\chi_{xxz}^{\rm ext}$. Explicitly,
\begin{eqnarray}
	\chi_{zyy}^{\rm ext}&=&\chi_{1}\int\frac{d^{3}k}{(2\pi)^{3}}(\partial_{k_{y}}f)\Omega_{x}\nonumber\\
	&=&-\chi_{1}\int\frac{d^{3}k}{(2\pi)^{3}}v_{y}(\bk)\delta(\mu-d(\bk))\Omega_{x}=-\chi_{yyz}^{\rm ext},\label{tensor1}\\
	\chi_{xxz}^{\rm ext}&=&\chi_{1}\int\frac{d^{3}k}{(2\pi)^{3}}(\partial_{k_{x}}f)\Omega_{y}\nonumber\\
	&=&-\chi_{1}\int\frac{d^{3}k}{(2\pi)^{3}}v_{x}(\bk)\delta(\mu-d(\bk))\Omega_{y}=-\chi_{zxx}^{\rm ext}.\label{tensor2}
\end{eqnarray}
Taking $v_{y}=2nv^{n}k_{\rho}^{2n-1}\sin\theta$ into Eq.(\ref{tensor1}), a straightforward calculation
gives
\begin{eqnarray}
	\chi_{zyy}^{\rm ext}&=&-\chi_{1}\int\frac{d^{3}k}{(2\pi)^{3}}\frac{(2n)^{2}v_{z}v^{4n}k_{\rho}^{4n-2}\sin\theta
		(m\sin\theta+v_{z}k_{z}\cos\theta)}{d^{2}(\bk)}\delta(\mu-d(\bk))\nonumber\\
	&=&-\frac{\chi_{1}}{8\pi^{3}}\int_{0}^{+\infty}k_{\rho}dk_{\rho}\int_{0}^{2\pi}d\theta\int_{-\infty}^{\infty}dk_{z}
	\frac{(2n)^{2}v_{z}v^{4n}k_{\rho}^{4n-2}\sin\theta
		(m\sin\theta+v_{z}k_{z}\cos\theta)}{d^{2}(\bk)}\delta(\mu-d(\bk))\nonumber\\
	&=&-\frac{\chi_{1}}{8\pi^{2}}\int_{0}^{+\infty}k_{\rho}dk_{\rho}\int_{-\infty}^{\infty}dk_{z}\frac{(2n)^{2}
		v_{z}mv^{4n}k_{\rho}^{4n-2}}{d^{2}(\bk)}\delta(\mu-d(\bk))\nonumber\\
	&=&-\frac{\chi_{1}}{8\pi^{2}}\int_{0}^{(\mu-m^{2})^{\frac{1}{2n}}/v}\frac{(2n)^{2}
		\text{sgn}(v_{z})mv^{4n}k_{\rho}^{4n-2}}{\mu^{2}\sqrt{\mu-m^2-v^{2n}k_{\rho}^{2n}}}k_{\rho}dk_{\rho}\Theta(\mu-m^{2})\nonumber\\
	&=&-\frac{n\chi_{1}\text{sgn}(v_{z})m}{2\mu^{2}\pi^{2}}\int_{0}^{\sqrt{\mu-m^{2}}}\frac{
		x^{3}}{\sqrt{\mu-m^2-x^{2}}}dx\Theta(\mu-m^{2})\nonumber\\
	&=&-\frac{n\chi_{1}\text{sgn}(v_{z})m(\mu-m^{2})^{3/2}}{2\mu^{2}\pi^{2}}\int_{0}^{\pi/2}
	\frac{\sin^{3}\phi}{\cos\phi}d\sin\phi\Theta(\mu-m^{2})\nonumber\\
	&=&-\frac{n\chi_{1}\text{sgn}(v_{z})m(\mu-m^{2})^{3/2}}{2\mu^{2}\pi^{2}}\int_{0}^{\pi/2}
	\sin^{3}\phi d\phi\Theta(\mu-m^{2})\nonumber\\
	&=&-\frac{n\chi_{1}\text{sgn}(v_{z}m)|m|(\mu-m^{2})^{3/2}}{3\mu^{2}\pi^{2}}\Theta(\mu-m^{2})\nonumber\\
	&=&N_{h}\chi_{1}\mathcal{B}_{1}(\mu,|m|)=N_{h}\chi_{1}\mathcal{B}_{1}(\frac{\mu}{m^{2}}),
\end{eqnarray}
where $\mathcal{B}_{1}(x)=\frac{2(x-1)^{3/2}}{3\pi^2 x^2}\Theta(x-1)$. Similar calculation gives $\chi_{zxx}^{\rm ext}=\chi_{zyy}^{\rm ext}$,
which is expected due to the existence of rotation symmetry.

\subsection{V. Intrinsic nonlinear Hall tensor for the low-energy Hopf Hamiltonian}

Without loss of generality, we again focus on the regime with
$\mu>0$.
The intrinsical nonlinear Hall conductivity tensor is given by~\cite{Gao2014INHE,Wang2021INHE,Liu2021INHE}
\begin{eqnarray}
	\chi_{\alpha\beta\gamma}^{\rm int}=-e^{3}\int \frac{d^{3}k}{(2\pi)^{3}}[\frac{\partial d(\bk)}{\partial k_{\alpha}}G^{(+)}_{\beta\gamma}(\bk)-\frac{\partial d(\bk)}{\partial k_{\beta}}G^{(+)}_{\alpha\gamma}(\bk)]\delta(\mu-d(\bk))
\end{eqnarray}
As the tensor is odd under the exchange of the first two subscripts, $\chi_{\alpha\alpha\gamma}^{\rm int}$
identically vanishes. On the other hand, according to the momentum dependence of all nonzero components of the BCP
tensor, one can find that $\chi_{\alpha\beta\beta}^{\rm int}$ identically vanishes since
the corresponding integrand is always odd under certain mirror operation. 

The above analysis indicates that only when $\alpha\neq\beta\neq\gamma$, the components of the intrinsic nonlinear Hall conductivity
tensor can be finite.  By straightforward calculations, we find that
\begin{eqnarray}
	\chi_{xyz}^{\rm int}&=&-e^{3}\int \frac{d^{3}k}{(2\pi)^{3}}[\frac{\partial d(\bk)}{\partial k_{x}}G^{(+)}_{yz}(\bk)-\frac{\partial d(\bk)}{\partial k_{y}}G^{(+)}_{xz}(\bk)]\delta(\mu-d(\bk))\nonumber\\
	&=&-e^{3}\int \frac{d^{3}k}{(2\pi)^{3}}\frac{2n^2v_{z}mv^{4n}k_{\rho}^{4n-2}}{d^{3}(\bk)}\delta(\mu-d(\bk))\nonumber\\
	&=&-n\frac{e^{3}\text{sgn}(v_{z})m(\mu-m^{2})^{3/2}}{3\pi^{2}\mu^{3}}\Theta(\mu-m^{2})\nonumber\\
	&=&N_{h}\frac{e^{3}}{m^{2}}\frac{2|m|^{3}(\mu-m^{2})^{3/2}}{3\pi^{2}\mu^{3}}\Theta(\mu-m^{2})
	=N_{h}\chi_{2}\mathcal{B}_{2}(\frac{\mu}{m^{2}}),
\end{eqnarray}
where $\chi_{2}=e^{3}/m^{2}$ and $\mathcal{B}_{2}(x)=(2(x-1)^{3/2}/3\pi^{2}x^{3})\Theta(x-1)$. Similar
calculations reveal
\begin{eqnarray}
	\chi_{yzx}^{\rm int}&=&-e^{3}\int \frac{d^{3}k}{(2\pi)^{3}}[\frac{\partial d(\bk)}{\partial k_{y}}G^{(+)}_{zx}(\bk)-\frac{\partial d}{\partial k_{z}}G^{(+)}_{xy}(\bk)]\delta(\mu-d(\bk))\nonumber\\
	&=&e^{3}\int \frac{d^{3}k}{(2\pi)^{3}}\frac{2n^2v_{z}mv^{4n}k_{\rho}^{4n-2}\sin^{2}\theta}{d^{3}}\delta(\mu-d(\bk))\nonumber\\
	&=&n\frac{e^{3}\text{sgn}(v_{z})m(\mu-m^{2})^{3/2}}{6\mu^{3}\pi^{2}}\Theta(\mu-m^{2})=-\frac{1}{2}\chi_{xyz}^{\rm int}.
\end{eqnarray}
and 
\begin{eqnarray}
	\chi_{zxy}^{\rm int}&=&-e^{3}\int \frac{d^{3}k}{(2\pi)^{3}}[\frac{\partial d(\bk)}{\partial k_{z}}G^{(+)}_{xy}(\bk)-\frac{\partial d}{\partial k_{x}}G^{(+)}_{zy}(\bk)]\delta(\mu-d(\bk))\nonumber\\
	&=&e^{3}\int \frac{d^{3}k}{(2\pi)^{3}}\frac{2n^2v_{z}mv^{4n}k_{\rho}^{4n-2}\cos^{2}\theta}{d^{3}}\delta(\mu-d(\bk))\nonumber\\
	&=&n\frac{e^{3}\text{sgn}(v_{z})m(\mu-m^{2})^{3/2}}{6\mu^{3}\pi^{2}}\Theta(\mu-m^{2})=-\frac{1}{2}\chi_{xyz}^{\rm int}.
\end{eqnarray}

\subsection{VI. Extrinsic and intrinsic Hall conductivity tensors across three-band Berry-dipole transitions}

Following Ref.~\cite{Graf2022hopf}, we consider a three-band model of the form
\begin{eqnarray}
	H(\bk)=\left(
	\begin{array}{ccc}
		0 & v^{n}(k_{x}+ik_{y})^{n} & m-iv_{z}k_{z} \\
		v^{n}(k_{x}-ik_{y})^{n} & 0 & 0 \\
		m+iv_{z}k_{z} & 0 & 0 \\
	\end{array}
	\right).
\end{eqnarray}
Rotation symmetry about the $z$ axis is also assumed.
The three bands of this Hamiltonian have the form
\begin{eqnarray}
	E_{\pm}(\bk)=\pm\sqrt{v^{2n}k_{\rho}^{2n}+v_{z}^{2}k_{z}^{2}+m^{2}},\quad E_{0}(\bk)=0.
\end{eqnarray}
At the critical point $m=0$, the three bands touch with each other and form a three-fold
nodal point.

The eigenstates of the Hamiltonian have the form
\begin{eqnarray}
	|u_{\pm}(\bk)\rangle&=&\frac{1}{\sqrt{2}E_{+}(\bk)}\left(
	\begin{array}{c}
		\pm E_{+}(\bk) \\
		v^{n}(k_{x}-ik_{y})^{n} \\
		m+iv_{z}k_{z} \\
	\end{array}
	\right),
	|u_{0}(\bk)\rangle=\frac{1}{E_{+}(\bk)}\left(
	\begin{array}{c}
		0 \\
		m-iv_{z}k_{z}  \\
		-v^{n}(k_{x}+ik_{y})^{n} \\
	\end{array}
	\right),
\end{eqnarray}
Similar to the main text, we consider $\mu>0$, so we only need
to calculate the Berry curvature and BCP of the top band.

The Berry curvature of the top band is given by
\begin{eqnarray}
	\Omega_{\alpha\beta}^{(+)}(\bk)=-2\text{Im}\langle \partial_{k_{\alpha}}u_{+}(\bk)|\partial_{k_{\beta}}u_{+}(\bk)\rangle.
\end{eqnarray}
Using this formula, one can find that
\begin{eqnarray}
	\Omega_{x}^{(+)}(\bk)\equiv\Omega_{yz}^{(+)}(\bk)
	&=&\frac{nv^{2n}k_{\rho}^{2(n-1)}k_{x}\partial_{k_{z}}E_{+}+mv_{z}\partial_{k_{y}}E_{+}}{E_{+}^{3}(\bk)}\nonumber\\
	&=&\frac{nv_{z}v^{2n}k_{\rho}^{2n-1}(v_{z}k_{z}\cos\theta+m\sin\theta)}{E_{+}^{4}(\bk)},\nonumber\\
	\Omega_{y}^{(+)}(\bk)\equiv\Omega_{zx}^{(+)}(\bk)
	&=&\frac{nv^{2n}k_{\rho}^{2(n-1)}k_{y}\partial_{k_{z}}E_{+}-mv_{z}\partial_{k_{x}}E_{+}}{E_{+}^{3}(\bk)}\nonumber\\
	&=&\frac{nv_{z}v^{2n}k_{\rho}^{2n-1}(v_{z}k_{z}\sin\theta-m\cos\theta)}{E_{+}^{4}(\bk)},\nonumber\\
	\Omega_{z}^{(+)}(\bk)\equiv\Omega_{xy}^{(+)}(\bk)
	&=&\frac{n^{2}v^{2n}k_{\rho}^{2(n-1)}}{E_{+}^{2}(\bk)}-\frac{nv^{n}k_{\rho}^{2(n-1)}}{E_{+}^{3}(\bk)}
	(k_{x}\partial_{k_{x}}E_{+}+k_{y}\partial_{k_{y}}E_{+})\nonumber\\
	&=&\frac{n^{2}v^{2n}k_{\rho}^{2(n-1)}(v_{z}^{2}k_{z}^{2}+m^{2})}{E_{+}^{4}(\bk)}.
\end{eqnarray}
Compared to the Berry curvature for the two-band Hopf model, one can find that they only differ
by a factor $2$. Thus, an integral of the Berry curvature over a closed surface enclosing the critical point also gives 
zero Chern number, but  an integral of the Berry curvature over the upper or lower half of the surface also
gives a quantized value. 

Since the Berry curvature only differs by a factor, the nonzero components of the extrinsic nonlinear Hall conductivity tensor 
are also $\chi_{zxx}^{\rm ext}$, $\chi_{zyy}^{\rm ext}$, $\chi_{xxz}^{\rm ext}$ and $\chi_{yyz}^{\rm ext}$. 
By straightforward calculations, one has 
\begin{eqnarray}
	\chi_{zxx}^{\rm ext}&=&\chi_{1}\int\frac{d^{3}k}{(2\pi)^{3}}v_{x}\Omega_{y}^{(+)}(\bk)\delta(\mu-E_{+}(\bk))\nonumber\\
	&=&
	\chi_{1}\int\frac{d^{3}k}{(2\pi)^{3}}\frac{n^{2}v_{z}v^{4n}k_{\rho}^{4n-2}\cos\theta(v_{z}k_{z}\sin\theta-m\cos\theta)}
	{E_{+}^{5}(\bk)}\delta(\mu-E_{+}(\bk))\nonumber\\
	&=&\frac{\chi_{1}}{8\pi^{3}}\int_{0}^{+\infty}k_{\rho}dk_{\rho}\int_{0}^{2\pi}d\theta\int_{-\infty}^{\infty}dk_{z}
	\frac{n^{2}v_{z}v^{4n}k_{\rho}^{4n-2}\cos\theta(v_{z}k_{z}\sin\theta-m\cos\theta)}
	{E_{+}^{5}(\bk)}\delta(\mu-E_{+}(\bk))\nonumber\\
	&=&-\frac{\chi_{1}}{8\pi^{2}}\int_{0}^{+\infty}k_{\rho}dk_{\rho}\int_{-\infty}^{\infty}dk_{z}
	\frac{n^{2}v_{z}mv^{4n}k_{\rho}^{4n-2}}
	{E_{+}^{5}(\bk)}\delta(\mu-E_{+}(\bk))\nonumber\\
	&=&-\frac{\chi_{1}}{4\pi^{2}}\int_{0}^{(\mu^{2}-m^{2})^{\frac{1}{2n}}/v}\frac{n^{2}
		\text{sgn}(v_{z})mv^{4n}k_{\rho}^{4n-2}}{\mu^{4}\sqrt{\mu^{2}-m^2-v^{2n}k_{\rho}^{2n}}}k_{\rho}dk_{\rho}\Theta(\mu-|m|)\nonumber\\
	&=&-\frac{n\chi_{1}\text{sgn}(v_{z})m}{4\mu^{4}\pi^{2}}\int_{0}^{\sqrt{\mu^{2}-m^{2}}}\frac{
		x^{3}}{\sqrt{\mu^{2}-m^2-x^{2}}}dx\Theta(\mu-|m|)\nonumber\\
	&=&-\frac{n\chi_{1}\text{sgn}(v_{z})m(\mu^{2}-m^{2})^{3/2}}{4\mu^{4}\pi^{2}}\int_{0}^{\pi/2}
	\frac{\sin^{3}\phi}{\cos\phi}d\sin\phi\Theta(\mu-|m|)\nonumber\\
	&=&-\frac{n\chi_{1}\text{sgn}(v_{z})m(\mu^{2}-m^{2})^{3/2}}{4\mu^{4}\pi^{2}}\int_{0}^{\pi/2}
	\sin^{3}\phi d\phi\Theta(\mu-|m|)\nonumber\\
	&=&-\frac{n\chi_{1}\text{sgn}(v_{z}m)|m|(\mu^{2}-m^{2})^{3/2}}{6\mu^{4}\pi^{2}}\Theta(\mu-|m|)\nonumber\\
	&=&-\frac{n\text{sgn}(v_{z}m)}{2}\chi_{1}\mathcal{Z}_{1}(\mu,|m|)=-\frac{n\text{sgn}(v_{z}m)}{2}\chi_{1}\mathcal{Z}_{1}(\frac{\mu}{|m|}),
\end{eqnarray}
where 
\begin{eqnarray}
	\mathcal{Z}_{1}(x)=\frac{(x^{2}-1)^{3/2}}{3\pi^{2}x^{4}}\Theta(x-1).
\end{eqnarray}
Similar calculations find $\chi_{zyy}^{\rm ext}=\chi_{zxx}^{\rm ext}$. Compared 
to the results for the two-band Hopf model, one sees that 
the difference is only in the forms of the universal functions.

To calculate the BCP for the top band, we apply the formula~\cite{Gao2014INHE,Wang2021INHE,Liu2021INHE}
\begin{eqnarray}
	G^{(+)}_{\beta\gamma}(\bk)=2\text{Re}\sum_{m=0,-}\frac{A_{+m,\beta}(\bk)A_{m+,\gamma}(\bk)}{E_{+}(\bk)-E_{m}(\bk)},
\end{eqnarray}
where $A_{+m,\beta}(\bk)=i\langle u_{+}(\bk)|\partial_{k_{\beta}}u_{m}(\bk)\rangle$ is the interband Berry connection.
By straightforward calculations, one finds that the interband Berry connection involving the top and middle bands have
the form
\begin{eqnarray}
	A_{+0,x}(\bk)&=&-i\frac{nv^{n}(k_{x}+ik_{y})^{n-1}(m-iv_{z}k_{z})}{\sqrt{2}E_{+}^{2}(\bk)},\nonumber\\
	A_{+0,y}(\bk)&=&\frac{nv^{n}(k_{x}+ik_{y})^{n-1}(m-iv_{z}k_{z})}{\sqrt{2}E_{+}^{2}(\bk)},\nonumber\\
	A_{+0,z}(\bk)&=&\frac{v_{z}v^{n}(k_{x}+ik_{y})^{n}}{\sqrt{2}E_{+}^{2}(\bk)},
\end{eqnarray}
and the interband Berry connections involving the top and bottom bands have
the form
\begin{eqnarray}
	A_{+-,x}(\bk)&=&-\frac{nv^{2n}k_{\rho}^{2n-1}\sin\theta}{2E_{+}^{2}(\bk)},\nonumber\\
	A_{+-,y}(\bk)&=&\frac{nv^{2n}k_{\rho}^{2n-1}\cos\theta}{2E_{+}^{2}(\bk)},\nonumber\\
	A_{+-,z}(\bk)&=&-\frac{mv_{z}}{2E_{+}^{2}(\bk)}.
\end{eqnarray}
Using the relation $A_{nm,\alpha}=A_{mn,\alpha}^{*}$, one finds that 
\begin{eqnarray}
	G^{(+)}_{xx}(\bk)&=&\frac{n^{2}v^{2n}k_{\rho}^{2(n-1)}(m^{2}+v_{z}^{2}k_{z}^{2})}{E_{+}^{5}(\bk)}+
	\frac{n^{2}v^{4n}k_{\rho}^{4n-2}\sin^{2}\theta}{4E_{+}^{5}(\bk)},\nonumber\\
	G^{(+)}_{yy}(\bk)&=&\frac{n^{2}v^{2n}k_{\rho}^{2(n-1)}(m^{2}+v_{z}^{2}k_{z}^{2})}{E_{+}^{5}(\bk)}+
	\frac{n^{2}v^{4n}k_{\rho}^{4n-2}\cos^{2}\theta}{4E_{+}^{5}(\bk)},\nonumber\\
	G^{(+)}_{zz}(\bk)&=&\frac{v_{z}^{2}v^{2n}k_{\rho}^{2n}}{E_{+}^{5}(\bk)}+
	\frac{m^{2}v_{z}^{2}}{4E_{+}^{5}(\bk)},\nonumber\\
	G^{(+)}_{xy}(\bk)&=&-\frac{n^{2}v^{4n}k_{\rho}^{4n-2}\cos\theta \sin\theta}{4E_{+}^{5}(\bk)},\nonumber\\
	G^{(+)}_{xz}(\bk)&=&-\frac{nv_{z}v^{2n}k_{\rho}^{2n-1}(m\sin\theta+v_{z}k_{z}\cos\theta)}{E_{+}^{5}(\bk)}
	+\frac{nv_{z}v^{2n}k_{\rho}^{2n-1}m\sin\theta}{4E_{+}^{5}(\bk)},\nonumber\\
	G^{(+)}_{yz}(\bk)&=&-\frac{nv_{z}v^{2n}k_{\rho}^{2n-1}(v_{z}k_{z}\sin\theta-m\cos\theta)}{E_{+}^{5}(\bk)}
	-\frac{nv_{z}v^{2n}k_{\rho}^{2n-1}m\cos\theta}{4E_{+}^{5}(\bk)}. 
\end{eqnarray}
Because of the two-part contributions, the BCP no longer has a simple relation with the Berry curvature as 
exhibited in the two-band Hopf model.  
However, as will be shown below, the extrinsic and intrinsic nonlinear Hall conductivity tensors still hold a simple relation. 
According to the momentum dependence of $G^{(+)}_{\alpha\beta}$, one can find that 
the nonzero components of the intrinsic nonlinear Hall conductivity tensor 
require the three subscripts of $\chi^{\rm int}_{\alpha\beta\gamma}$ all to be different, 
i.e., $\alpha\neq\beta\neq\gamma$. By straightforward calculations, one has 
\begin{eqnarray}
	\chi^{\rm int}_{xyz}=&=&-e^{3}\int \frac{d^{3}k}{(2\pi)^{3}}[\frac{\partial E_{+}(\bk)}{\partial k_{x}}G^{(+)}_{yz}(\bk)-\frac{\partial E_{+}(\bk)}{\partial k_{y}}G^{(+)}_{xz}(\bk)]\delta(\mu-E_{+}(\bk))\nonumber\\
	&=&-e^{3}\int \frac{d^{3}k}{(2\pi)^{3}}\frac{3n^2v_{z}mv^{4n}k_{\rho}^{4n-2}}{4E^{6}_{+}(\bk)}\delta(\mu-E_{+}(\bk))\nonumber\\
	&=&-\frac{3e^{3}}{16\pi^{2}}\int_{0}^{+\infty}k_{\rho}dk_{\rho}\int_{-\infty}^{+\infty}dk_{z}
	\frac{n^2v_{z}mv^{4n}k_{\rho}^{4n-2}}{E^{6}_{+}(\bk)}\delta(\mu-E_{+}(\bk)),\nonumber\\
	&=&-\frac{3n\text{sgn}(v)e^{3}}{8\pi^{2}}\int_{0}^{(\mu^{2}-m^{2})^{\frac{1}{2n}}/v}
	\frac{nmv^{4n}k_{\rho}^{4n-2}}{\mu^{5}\sqrt{\mu^{2}-m^{2}-v^{2n}k_{\rho}^{2n}}}\Theta(\mu-|m|)k_{\rho}dk_{\rho},\nonumber\\
	&=&-\frac{3n\text{sgn}(v)me^{3}}{8\pi^{2}\mu^{5}}\Theta(\mu-|m|)\int_{0}^{\sqrt{\mu^{2}-m^{2}}}
	\frac{x^{3}}{\sqrt{\mu^{2}-m^{2}-x^{2}}}dx,\nonumber\\
	&=&-n\frac{\text{sgn}(v_{z})me^{3}(\mu^{2}-m^{2})^{3/2}}{4\pi^{2}\mu^{5}}\Theta(\mu-|m|)\nonumber\\
	&=&-\frac{n\text{sgn}(v_{z}m)}{2}\frac{e^{3}}{|m|}\frac{|m|^{2}(\mu^{2}-m^{2})^{3/2}}{2\pi^{2}\mu^{5}}\Theta(\mu-|m|)
	=-\frac{n\text{sgn}(v_{z}m)}{2}\tilde{\chi}_{2}\mathcal{Z}_{2}(\frac{\mu}{|m|}),
\end{eqnarray}
where $\tilde{\chi}_{2}=e^{3}/|m|$, and the universal function $\mathcal{Z}_{2}(x)$ has the form 
\begin{eqnarray}
	\mathcal{Z}_{2}(x)=\frac{(x^{2}-1)^{3/2}}{2\pi^{2}x^{5}}\Theta(x-1)=\frac{2\mathcal{Z}_{1}(x)}{3x}.
\end{eqnarray}
It is worth noting that while $\tilde{\chi}_{2}$ takes a form somewhat different from $\chi_{2}=e^{3}/m^{2}$
appearing in the two-band Hopf model, they are in fact equivalent since $|m|$ 
has the physical meaning as energy gap for the three-band model considered. 

Similar calculations give 
\begin{eqnarray}
	\chi_{yzx}^{\rm int}&=&-e^{3}\int \frac{d^{3}k}{(2\pi)^{3}}[\frac{\partial E_{+}(\bk)}{\partial k_{y}}G^{(+)}_{zx}(\bk)-\frac{\partial E_{+}}{\partial k_{z}}G^{(+)}_{xy}(\bk)]\delta(\mu-E_{+}(\bk))\nonumber\\
	&=&e^{3}\int \frac{d^{3}k}{(2\pi)^{3}}\frac{3n^2v_{z}mv^{4n}k_{\rho}^{4n-2}(m\sin\theta+v_{z}k_{z}\cos\theta)\sin\theta}{4E_{+}^{6}(\bk)}\delta(\mu-E_{+}(\bk))\nonumber\\
	&=&\frac{3e^{3}}{32\pi^{2}}\int_{0}^{+\infty}k_{\rho}dk_{\rho}\int_{-\infty}^{+\infty}dk_{z} \frac{n^2v_{z}mv^{4n}k_{\rho}^{4n-2}m}{E_{+}^{6}(\bk)}\delta(\mu-E_{+}(\bk))\nonumber\\
	&=&-\frac{1}{2}\chi_{xyz}^{\rm int}.
\end{eqnarray}
and 
\begin{eqnarray}
	\chi_{zxy}^{\rm int}&=&-e^{3}\int \frac{d^{3}k}{(2\pi)^{3}}[\frac{\partial E_{+}(\bk)}{\partial k_{z}}G^{(+)}_{xy}(\bk)-\frac{\partial E_{+}}{\partial k_{x}}G^{(+)}_{zy}(\bk)]\delta(\mu-E_{+}(\bk))\nonumber\\
	&=&e^{3}\int \frac{d^{3}k}{(2\pi)^{3}}\frac{3n^2v_{z}v^{4n}k_{\rho}^{4n-2}(m\cos\theta-v_{z}k_{z}\sin\theta)\cos\theta}{4E_{+}^{6}(\bk)}\delta(\mu-d(\bk))\nonumber\\
	&=&\frac{3e^{3}}{32\pi^{2}}\int_{0}^{+\infty}k_{\rho}dk_{\rho}\int_{-\infty}^{+\infty}dk_{z} \frac{n^2v_{z}mv^{4n}k_{\rho}^{4n-2}m}{E_{+}^{6}(\bk)}\delta(\mu-E_{+}(\bk))\nonumber\\
	&=&-\frac{1}{2}\chi_{xyz}^{\rm int}.
\end{eqnarray}
It is readily seen that the components of the intrinsic Hall conductivity tensor hold the same relation as 
the two-band Hopf model. 

From the results presented above, one sees that the extrinsic and intrinsic Hall conductivity tensors 
also display a general-sense quantized behavior across the three-band Berry-dipole transitions. The only difference 
is that the universal functions are different to those for the two-band Hopf model. 

\end{widetext}

\end{document}